\documentstyle[12pt,epsfig]{article}
\def\Bll{$B \rightarrow X_s l^+ l^-$}
\def\BLL{$\overline{B} \rightarrow \overline{X}_s l^+ l^-$}
\def\Bee{$B \rightarrow X_s e^+ e^-$}
\def\Bmm{$B \rightarrow X_s \mu^+ \mu^-$}
\def\Btt{$B \rightarrow X_s \tau^+ \tau^-$}

\def\Bsg{B \rightarrow X_s \gamma}
\def\phm{$\phi_\mu$}
\def\pha{$\phi_{A_0}$}
\def\beq{\begin{eqnarray}}
\def\eeq{\end{eqnarray}}
\def\nnb{\nonumber}

\begin{document}
\bigskip
{\large\bf
\centerline{The Electric Dipole Moment and CP Violation in \Bll }
\centerline{ in
SUGRA Models with Nonuniversal Gaugino Masses}
\bigskip
\normalsize

\centerline{Chao-Shang Huang,~~~Liao Wei}
\centerline{\sl Institute of Theoretical Physics, Academia Sinica,
      P.O.Box 2735,}
\centerline{\sl Beijing 100080,P.R.China}
\bigskip

\begin{abstract}
  The constraints of electric dipole moments (EDMs) of electron and neutron
on the parameter space in supergravity (SUGRA) models with nonuniversal 
gaugino masses are analyzed. It is shown that with a light sparticle spectrum
, the sufficient cancellations in the calculation of EDMs can happen for
all phases being order of one in the small tan$\beta$ case and all phases
but $\phi_{\mu}$ ($|\phi_{\mu}| \stackrel{<}{\sim} \pi/6$) order of one in 
the large tan$\beta$ case. This is in contrast to the case of mSUGRA in
which  in the parameter space where cancellations among various SUSY
contributions to EDMs happen $|\phi_{\mu}|$ must be less than $\pi/10$
for small $tan\beta$ and ${\cal{O}}(10^{-2})$ for large $tan\beta$.
Direct CP asymmetries and the T-odd normal polarization of lepton
in $B\rightarrow X_s l^+l^-$ are investigated in the models. In the large 
tan$\beta$ case,  $A_{CP}^2$ and  $P_N$ for l=$\mu$ ( $\tau$) can be enhanced
by about a factor of  ten ( ten) and ten (three) respectively compared to
those of mSUGRA.
\end{abstract}

\newpage

~~~~Recent observation of Re($\epsilon'$/$\epsilon$) by KTev collaboration
\cite{kt} definitely confirms the earlier NA31 experiment\cite{na}. This
direct CP violation measurement in the Kaon system can be accommodated by
the CKM phase in standard model (SM) within the theoretical uncertainties.
However, the CKM phase is not enough to explain the matter-antimatter asymmetry 
in the universe and gives the contribution to EDMs much smaller than the limits 
of EDMs of electron and neutron. One needs to
have new sources of CP violation and examine their phenomenological effects.

There exist new sources of CP violation in SUSY theories which come from the
phases of soft SUSY breaking parameters. It is well-known for a long time
that in order to satisfy the current experimental limits on EDMs of electron
and neutron SUSY CP-violating phases have to be much smaller($ \stackrel{<}
{\sim} 10^{-2}$) unless sfermion masses of the first and second generations
are very large ($>$ 1 Tev)~\cite{old}. Recently it has been shown that various
contributions to EDMs cancel with each other in significant regions of the 
parameter space so that the current experimental limits on the EDM of 
electron (EDME)~\cite{commins} and the EDM of neutron (EDMN)~\cite{HLG},
\beq
|d_e| < 4.3\times 10^{-27} ecm
\eeq
and
\beq
|d_n| < 6.3\times 10^{-26} ecm ,
\eeq
can be satisfied for SUSY models with SUSY phases of order one and relatively
light sparticle spectrum ($<$ 1 Tev)~\cite{IN,others}. In mSUGRA even in the
parameter space where cancellation among various SUSY contributions to
neutron EDM(EDMN) happens $|\phi_{\mu}|$ must be less than $\pi/10$ for small
$tan\beta$\cite{IN} and ${\cal{O}}(10^{-2})$ for large $tan\beta$~\cite{HL}
while the allowed range of $\phi_{A_0}$ is almost unconstrained. Brhlik et al. 
pointed out that more sufficient cancellations happen in MSSM if gaugino 
masses are complex~\cite{bgk}. In the letter we consider cancellation
phenomena in SUGRA with nonuniversal gaugino masses.

CP violation has so far only been observed in K system. It is one of the goals
of the B factories presently under construction to discover and examine CP
violation in the B system. Direct CP violation in \Bll ~in SM has been examined
and the result is that it is unobservablly small~\cite{KSAH}. In mSUGRA the
CP assymmetry of branching ratio on \Bll has been given in~\cite{goto}.
A detailed analysis of SUSY contributions to CP Violation in semileptonic
B decays has been performed using the mass insertion approximation in~\cite{ls}.
Direct CP asymmetries and the T-odd normal polarization of lepton
in \Bll ~in mSUGRA with CP-violating phases are investigated in our 
previous paper~\cite{HL}. In the letter we extend the investigation to 
SUGRA models with nonuniversal gaugino masses after studying the allowed
regions of the parameter space in the models by EDM data.

In order to concentrate on the effects of the phases arising from complex 
gaugino masses we limit ourself to a class of SUGRA models with nonuniversal
gaugino masses in which gaugino masses at high energy scale (GUT scale) are
nonuniversal but scalar masses and trilinear couplings at GUT scale are still
universal. Such a class of effective SUGRA models can naturally arise from
string models~\cite{bim}. In this class of models, compared to the mSUGRA,
there are two more new independent phases~\cite{bgk} which we choose to be
$\phi_1$ and $\phi_3$, the phases of gaugino masses $M_1$ and $M_3$, in
addition to the phases $\phi_{\mu}$ and $\phi_{A_0}$. From the one loop
renormalization group equations (RGEs) for $M_i$ (i=1,2,3)~\cite{rge}
\beq
\frac{dM_i}{dt} = \frac{1}{4 \pi} b_i \alpha_i M_i   ~~~~~~~i=1,2,3
\eeq
where $\alpha_i=\frac{g^2_i}{4 \pi}$, $t=ln(Q^2/M_{GUT}^2)$, we know that 
the phases of $M_i$ do not run, like the phase of $\mu$.

Let us recall the cancellation mechanism for EDME. There are only two 
contributions, the chargino (-sfermion loop) and neutrilino (-sfermion loop)
contributions, to the EDME. The chargino contribution involves gaugino
-Higgsino(g-h) mixing. The neutrilino contribution involves not only 
gaugino-Higgsino mixing but also gaugino-gaugino(g-g) mixing. The chargino 
contribution and the part of the neutrilino contribution which comes from
g-h mixing have automatically opposite sign because of the opposite sign of
$\mu$ in chargino and neutrilino mass matrices. In general, the chargino 
contribution in magnitude is significantly larger than the part of neutrilino
contribution. Therefore, as pointed out in ref.~\cite{bgk}, a cancellation 
can happen only if the another part of the neutrilino contribution which
comes from the g-g mixing can balance some of the difference between the two 
contributions. For EDME the neutrilino contribution which comes from the g-g
mixing is proportional to~\cite{bgk}
\beq
\frac{1}{m^2_{\tilde{e}}} m_e [ A_e sin(\phi_{A_e}-\phi_1) + |\mu|
tan\beta sin(\phi_\mu +\phi_1)]
\eeq
Therefore, given \phm, the sign of the contribution can be controlled by 
choosing $\phi_1$ and $\phi_{A_0}$ and the magnitude of the contribution
can increase by increasing $\mu$ and/or $A_e$. Because the chargino 
contribution is dependent on $\mu$ and independent on $\phi_{A_e}$, it is
sufficient to have a cancellation that the magnitude of the $A_e$ term 
(first term) in eq.(4) is comparable to that of the $\mu tan\beta$ term
(second term) in eq.(4). This is easy to be down in MSSM in which $A_e$
and $\mu$ are free parameters. Thus, an almost exact cancellation can occur
for the whole range of $\phi_{\mu}$. That is exactly what happens in MSSM
~\cite{bgk}.

However, in SUGRA models low energy properties are determined by running 
RGEs from the high energy scale to the electroweak(EW) scale and the 
radiative breaking  mechanism of the EW symmetry puts constraints on CP-
violating phases. As long as we limit our discussion to mass spectra less than
than 1 Tev, $M_3$ and $A_0$(hence $A_e$) can not be too large. For small tan
$\beta$(say $\stackrel{<}{\sim} 2$), the sufficient condition (i.e., the two
terms in eq.(4) have size of the same order) can easily be realized in the
almost whole range of $\phi_{\mu}$ by choosing $\phi_{A_0}$ and $\phi_1$.
For moderate and large tan$\beta$, it is difficult for the condition to be
realized due to the limited values of $A_0$ (hence $A_e$) so that only
for some limited ranges of \phm ~the EDM constraint can be satisfied.
The similar (but more complicated) situation occurs for EDMN with 
appropriate $\phi_3$ as well as $\phi_{A_0}$ chosen.

In order to show the important role of $\phi_{A_0}$ played in the 
cancellation mechanism, in fig.1a and 1b we display EDME as function of
$\phi_1$ for $\phi_{A_0}$=0 and different \phm ~for both small tan$\beta$
(2) and large tan$\beta$ (30) cases. We can see from the fig.1 that most
of the range of $\phi_{\mu}$ is excluded by EDME in both cases. For EDMN as
function of $\phi_3$, similar results are obtained. That is, like $\phi_1$,
for positive \phm ~cancellations happen in some narrow ranges within
[$\pi$,$2 \pi$] of $\phi_3$ and within [0,$\pi$] for negative \phm. When we
vary the values of \pha ~we achieve the above mentioned results: almost whole
range of \phm ~is allowed by EDME and EDMN for small tan$\beta$ and
$|\phi_{\mu}| \stackrel{<}{\sim} \pi/6$ for large tan$\beta$(see fig.2).
Moreover, because $\phi_1 (\phi_3)$ is correlated with $\phi_{\mu}$, we
find that with varying \phm ~the whole range of $\phi_1$ and $\phi_3$ can
be allowed by EDM constraints. For large tan$\beta$ (30) case largest
$|\phi_{\mu}|$(about $\pi/6$) correspond to $\phi_1$ and $\phi_3$ around
$\mp \pi/2 \mp \pi/6$, while $\phi_1$ and $\phi_3$ are around $\pm \pi/2$
when \phm ~about $\mp 0.4$, and when \phm ~is about $\pm 0.2$ they are
around $\mp \pi/4$. The correlated values of $\phi_3$ and  \phm ~are
needed in analyses of $B\rightarrow X_s l^+l^-$ and $\Bsg$ below.
Correlation between \phm ~and tan$\beta$, with the absolute value of soft
breaking terms chosen as those in fig1 and appropriate phases chosen, 
is shown in fig.2 where all of the points 
are allowed by the experimental bounds on EDME and EDMN. One can see in 
the figure that \phm ~becomes more constrained as $tan\beta$ is increased.
Nevertherless, for $tan\beta$ larger than 6 the allowed regions of \phm
~are almost unchanged, which means that effects of the $A_e$ term (and $A_d$
term in the case of EDMN) in eq.(4) are relatively small and the balance 
is provided by the $\mu tan\beta$ term in eq.(4). Since we also consider 
the large $tan\beta$ case, we include the two loop contribution given by
D. Chang et.al\cite{ckp}. But the numerical calculations in the regions of 
the parameter space in which one loop EDMs satisfy the current experimental 
limits due to the cancellation mechanism show that it is very small compared 
to one loop contributions. 
 
We now turn to the calculations of the CP violation in \Bll. The direct CP
asymmetries in decay rate and backward-forward assymmetry for \Bll ~and \BLL
~are defined by~\cite{HL,hz}
\beq
&& A_{CP}\hspace{-12pt}^1 \hspace{12pt}(\hat{s}) = \frac{{\rm d} \Gamma
/{\rm d} \hat{s} - {\rm d}\overline{\Gamma} /{\rm d} \hat{s}}{{\rm d}
\Gamma /{\rm d} \hat{s} +{\rm d} \overline{\Gamma}/{\rm d} \hat{s}}
=\frac{D(\hat{s})-\overline{D}(\hat{s})}{D(\hat{s})+\overline{D}(\hat{s})} ,
\nnb \\ 
&& A_{CP}\hspace{-12pt}^2 \hspace{12pt} (\hat{s}) = \frac{A(\hat{s})-
\overline{A}(\hat{s})}{A(\hat{s})+\overline{A}(\hat{s})}
\eeq
where
\beq
&& A(\hat{s}) = 3 \sqrt{ 1-\frac{4 t^2}{\hat{s}}}\frac{E(\hat{s})}{D(\hat{s})}
, \nnb \\
&& D(\hat{s}) =  4 \big|C_7 \big|^2 (1+\frac{2}{\hat{s}}) (1+\frac
{2 t^2}{\hat{s}})+\big|{C_8}\hspace{-4pt}^{eff} \big|^2 (2 \hat{s}+1)
(1+\frac{2 t^2}{\hat{s}})+ \big|C_9 \big|^2 \big[ 1+2 \hat{s}
+ (1-4 \hat{s}) \frac{2 t^2}{\hat{s}} \big] \nonumber \\
&& +12 Re({C_8}\hspace{-4pt}^{eff} C_7\hspace{-4pt}^* ) (1+\frac{2 t^2}
{\hat{s}})+ \frac{3}{2} \big|C_{Q_1} \big|^2 (1-\frac{4 t^2}{\hat{s}})
\hat{s}+ \frac{3}{2} \big|C_{Q_2} \big|^2 \hat{s}+ 6 Re(C_9 C_{Q_2}^*) t \nnb \\
&& E(\hat{s}) = Re( C_8\hspace{-5pt}^{eff} C_9\hspace{-2pt}^* \hat{s}
+2 C_7 C_9 \hspace{-2pt}^* +C_8\hspace{-5pt}^{eff} C_{Q_1}\hspace{-5pt}^* t
 + 2 C_7 C_{Q_1}\hspace{-5pt}^* t)
\eeq

Another observable related to CP violating effects in \Bll ~is the normal 
polarization of the lepton in the decay, $P_N$, which is the T-odd
projection of the lepton spin onto the normal of the decay plane,
i.e $P_N  \sim \vec{s}_l\cdot (\vec{p}_s\times \vec{p}_{l^-})$~\cite{Lee}.
A straightforward calculation leads to\cite{HL,Rai}
\beq
  P_N = \frac{3 \pi}{4} \sqrt{1-\frac{4 t^2}{\hat{s}}} \hat{s}^\frac{1}{2}
Im\Big[2 C_8 \hspace{-2pt}^{eff*} C_9 t +4 C_9 C_7\hspace{-2pt}^*
\frac{t}{\hat{s}}+ C_8\hspace{-2pt}^{eff*} C_{Q_1}+ 2 C_7\hspace{-2pt}
^* C_{Q_1}+ C_9\hspace{-2pt}^* C_{Q_2} \Big] \Big/ D(\hat{s})
\eeq
The Wilson coefficients $C_i$ and $C_{Q_i}$ in eqs.(6) and (7) have been
given in ref.\cite{HL,hy,bb}. Since only $C_8^{eff}$ contains the non-trivial
strong phase, $A_{CP}^1$ is determined by Im$C_7$ and $A_{CP}^2$ by 
Im$C_{Q_1}$ and Im$C_7$. Although $P_N$ depends on all the relevant Wilson
coefficients a large $P_N$ does require relatively large values of Im$C_{Q_i}
(i=1,2)$~\cite{HL}. With the main contributions coming from exchanging
chargino-stop loop with neutral Higgs coupled to external b quark
\cite{hy}, imaginary parts of $C_{Q_i}$s come mainly from terms proportional
to (unitarity condition for stop mixing matrix has been used)
\beq
 \frac{m_{\chi_i} m_t}{m^2_W sin\beta ~cos\beta} U(i,2) V(i,2)
D_{t21} D^{*}_{t11}, ~~~~ i=1,2
\eeq
i.e CP violating phases enter into the imaginary parts of $C_{Q_i}$ through
g-h mixings (U, V) and chiral mixing ($D_t$) of stops. From the chargino mass
matrix 
\beq
M_C = \left(\matrix{M_2 & \sqrt{2} m_W  sin\beta \cr
        \sqrt{2} m_W cos\beta & \mu}
            \right)
\eeq
and stop mass matrix
\beq
M_{\tilde{t}}^2=\left(\matrix{M_{\tilde{Q}}^2+m{_t}^2+M_{z}^2(\frac{1}{2}-Q_u
\sin^2\theta_W)\cos2\beta & m_t(A_{t}^{*}-\mu \cot\beta) \cr
                        m_t(A_{t} -\mu^{*} \cot\beta) & M_{\tilde{U}
}^2+m{_t}^2+M_{z}^2 Q_u \sin^2\theta_W \cos2\beta}
                \right),
\eeq
we know that $\sum_{i=1}^2 m_{\chi_i} U(i,2) V(i,2) = \mu$ and $D_{t21}
D^{*}_{t11}=\frac{m_t}{m^2_{\tilde{t}_1} -m^2_{\tilde{t}_2}}(A_t -\mu^* 
cot\beta)$. Therefore, $A_t$ itself is as important
as $\mu$ for providing imaginary contributions to $C_{Q_i}$, in particular,
for large tan$\beta$. The similar conclusion holds also for $C_7$. 

We have known well that $A_t$ at the EW scale mainly depends on $M_3$ at the
GUT scale through RGE effects\cite{CKKM,aad}. In fact there exists a quasi
fixed point which shows the ratio of $A_t$ at $m_Z$ scale to $M_3$ at GUT
scale to be about $-1.6$ provided the Yukawa couplings of the third generation
are large enough~\cite{CKKM}. Hence it is possible for $A_t$ to achieve a
very large imaginary parts in the non-universal gaugino mass models, in 
contrast to the case of mSUGRA. Especially in large $tan\beta$ case, $|\phi_
{\mu}|$ is limited by EDM data to be less than $\pi/6$, so $A_t$ plays a more
important role in CP Violation than $\mu$.

To study the effect of large $\phi_3$ (hence large $\phi_{A_t}$) on $C_7$, 
we notice that essentially $A_t$ is multiplied by $\mu$. Changing the
sign of $A_t$ has the same effects of switching the sign of $\mu$ and
switching the sign of $\mu$ results in a sign change in $C_7$ (because
in most of the parameter space $\mu$ are much larger than the non-diagonal
terms in eq.(9)), so if $\phi_3$ ~is in [$\pi/2$, $3\pi /2$] (hence \pha
 ~in [$-\pi /2, \pi /2$]) and $\phi_{\mu}$ in [$-\pi /2, \pi /2$]), 
supersymmetry contributions give enhancement to $ReC_7$ so that it is hard
to satisfy the $\Bsg$ constraints: $ 2.\times 10^{-4}< Br(B \to X_s \gamma)
< 4.5 \times 10^{-4}$ \cite{cleo}. Similar situation occurs for $\phi_3$
being in [$-\pi /2, \pi /2$]) and \phm ~in [$\pi /2$,$3\pi /2$]. Since
supersymmetry gives large contributions to $C_7$ only when $tan\beta$ is 
large we shall focus on large $tan\beta$ case in the following. From the
above analysis of EDMN we know that for somewhat large \phm(say around $\mp
 0.4$) cancellations happen for $\phi_3$ near $\pm\pi/2$. With this kind of 
phases of $M_3$ and $\mu$ SUSY contributions to $C_7$ are almost totally
imaginary. So the value of Im$C_7$ is constrained to be very small by the 
branching ratio of $\Bsg$. One way to avoid the $\Bsg$ constraint is to
make use of the cancellation happened at $\phi_3$ around $\pm \pi/4$ and
\phm ~about $\mp 0.2$, With such kind of phases and a low mass spectrum
($|M_2|$ and $|M_3|$ around 150 GeV) the real 
part of SUSY contributions to $C_7$ cancels those from W-top and charged 
Higgs-top loops. The real part can even be cancelled to be near zero and
only a large imaginary part of $C_7$ remains. Another way is to just
suppress the total SUSY contributions to $C_7$, i.e., to make the mass spectrum
heavier but still less than 1 Tev ($|M_2|$ and $|M_3|$ larger than about 300 GeV
and less than about 500 GeV). In the former case  $A^1_{CP}$ can
reach order $1\%$ for \Bee. For \Bmm ~and ~\Btt, because of their larger
Yukawa coupling there are great enhancement of branching ratio\cite{hy} so
that $A^1_{CP}$ are smaller than that for \Bee. In the later case(we shall
call it as region A hereafter) it can only be a few thousandth at most, i.e.,
the same order as that in SM. 

As pointed out above, the effects of large $\phi_3$ on $C_{Q_i}$ are similar
to those on $C_7$. In large tan$\beta$ case, for $\phi_{\mu}\sim \pm 0.2$
and $\phi_3\sim \mp\pi/4$ or $\phi_{\mu}\sim \pi \pm 0.2$ and $\phi_3 \sim
\pi \mp \pi/4$ and with $|M_2|$ and $|M_3|$ around 150 GeV(which we shall
call as region B for simplicity), which are 
allowed by the EDME and EDMN limits, Im$C_{Q_i}$ reachs maxima. In small
tan$\beta$ case, although the constraints of EDMs on $\phi_{\mu}$ and
$\phi_3$ are relaxed the magnitude of Im$C_{Q_i}$ is very small since
$C_{Q_i}$ is proportional to $m_l tan^2\beta$(even $m_l tan^3\beta$ in some
regions of the parameter space). Therefore, we expect the significant CP 
violation in large tan$\beta$ case.

From eqs.(5) and (6),  $A_{CP}^2$ can be rewritten as
\beq
A_{CP}^2 = \frac{ E(\hat{s}) \overline{D}(\hat{s})- \overline{E}(\hat{s})
D(\hat{s})}{E(\hat{s}) \overline{D}(\hat{s})+\overline{E}(\hat{s})D(\hat{s})}
\nnb
\eeq
For l=e, the difference between $E(\hat{s})$ and $\overline{E}
(\hat{s})$ can be neglected(as it is proportional to lepton mass square,
see eq.(6)). So $A_{CP}^2$ for l=e can be reduced to 
\beq
A_{CP}^2 \doteq \frac{\overline{D}(\hat{s})- D(\hat{s})}{\overline{D}
(\hat{s})+ D(\hat{s})} \nnb
\eeq
that is exactly the opposite of $A_{CP}^1$. The same conclusion can be drawn 
for l=$\mu, \tau$ in small tan$\beta$ case due to smallness of $C_{Q_1}$.
On the other hand, for l=$\tau$ in large tan$\beta$ case,
$|E(\hat{s})-\overline{E}(\hat{s})|$ can be more important than $|D-\bar D|$
and consequently one has approximately
\beq
A_{CP}^2 \doteq \frac{ E(\hat{s})- \overline{E}(\hat{s})}{E(\hat{s})+
\overline{E}(\hat{s})} \nnb
\eeq
Thus, it is propotional to t~Im$C_{Q1}$. Therefore, in region B where 
$C_{Q_i}$s reach the maxima,
$A_{CP}^2$ can be over $50\%$. In region A, $C_{Q_i}$s are less important
and $A_{CP}^2$ can reach about $5\%$ at most. The correlation between 
$A_{CP}^2$ and EDME (or EDMN) in region A is plotted in fig.3 (note that we 
choose $\hat{s}=0.76$ as representative in the figure). For l=$\mu$,
the magnitude of $A_{CP}^2$ can be estimated to be of order $1\%$ at most
in the large tan$\beta$ case. Numerical calculations prove this estimate.

Fig.4 shows the correlation of EDM constraints and $P_N$ of \Btt in region
A. We can see in this figure that $P_N$ can reach more than 15 percent. In
region B, as $C_{Q_i}$s are much larger the numerator in eq.(7) is increased
a lot. But the denominator in eq.(7) (and consenquently the branching ratio 
of \Btt) is also greatly enhanced in this region, so $P_N$ is
just about $15\%$, i.e., not larger than the magnitude that can be achieved
in region A. Situations for muon are similar and because of its much
smaller Yukawa coupling the magnitude of $P_N$ can only reach about $6\%$.
For electron $P_N$ is negligibly small, due to its neglegibly small mass. 
An important feature that can be seen from fig.3 and fig.4 is that the 
magnitudes of $A_{CP}^2$ and $P_N$ will not be reduced if EDM constraints
improved. That is because the regions of parameter space where EDM 
constraints are satisfied are of width about $\pi/20$ for $\phi_3$
and about $\pi/4$ for \pha (adjustment needed), while $C_{Q_i}$s
do not change sharply within these regions. 

In summary, we have analyzed the constraints of electric dipole moments of 
electron and neutron on the parameter space in supergravity models with 
nonuniversal gaugino masses. It is shown that with a light sparticle spectrum,
the sufficient cancellations in the calculation of EDMs can happen due to
the presence of the two new phases arising from complex gaugino masses,
in addition to the phases $\phi_{\mu}$ and $\phi_{A_0}$. With appropriate 
correlation between $\phi_{\mu}$ and $\phi_1$ (for EDME) or $\phi_3$ (for
EDMN) as well as an appropriate choice of $\phi_{A_0}$, cancellations can
occur and all phases can be order of one in the small tan$\beta$ case and all
phases but $\phi_{\mu}$ ($|\phi_{\mu}| \stackrel{<}{\sim} \pi/6$) order of 
one in the large tan$\beta$ case. This is in contrast to the case of mSUGRA
where in the parameter space where cancellations among various SUSY
contributions to EDMs happen \phm ~must be less than $\pi/10$ for small 
$tan\beta$ and ${\cal{O}}(10^{-2})$ for large $tan\beta$. And our analysis
show that the branching ratio of $\Bsg$ gives an extra constraint on the
phases for large tan$\beta$ case with light mass spectrum. We have 
calculated direct CP asymmetries and the T-odd normal polarization of lepton
in \Bll ~in the regions of the parameter space in the models where the 
constraints from EDMs as well as $\Bsg$ are satisfied. It is shown that the
results for $A_{CP}^1$  are similar to those in mSUGRA if the mass sprectrum is
relatively heavier($|M_2| \& |M_3| \stackrel{>}{\sim} 300 GeV$) and it is also 
true for $A_{CP}^2$ in the small tan$\beta$ case. The former is due to the
constraint from $\Bsg$ and the latter is due to smallness of the contributions
from exchanging neutral Higgs bosons in the small tan$\beta$ case. However,
in the large tan$\beta$ case,  $A_{CP}^2$ can reach $1\%$ for $l=\mu$ and
is a few percent in most of allowed regions and can reach $50\%$ in some 
allowed regions for l=$\tau$. $A_{CP}^2 $ for $l=e$ is approximately equal 
to $A_{CP}^1$ even in the large tan$\beta$ case. $P_N$ can reach $6\%$ for
$l=\mu$ and is in the range from $1\%$ to $15\%$ in most of the allowed 
regions for l=$\tau$ in the large tan$\beta$ case. That is, there is a 
significant enhancement compared to the mSUGRA in which $P_N$ only can reach
about $0.5\%$ for l=$\mu$ and about $5\%$ for l=$\tau$. In the small 
tan$\beta$ case the results are similar to those in mSUGRA.
For l=e, $P_N$ is negligibly small, as it should be. \\

This research was supported in part by the National Natural Science 
Foundation of China. Thanks to T. Nath and A. Pilaftsi for their
helpful communications.

\newpage
\begin{figure}[t]
\vspace{-2cm}
\centerline{
\epsfig{figure=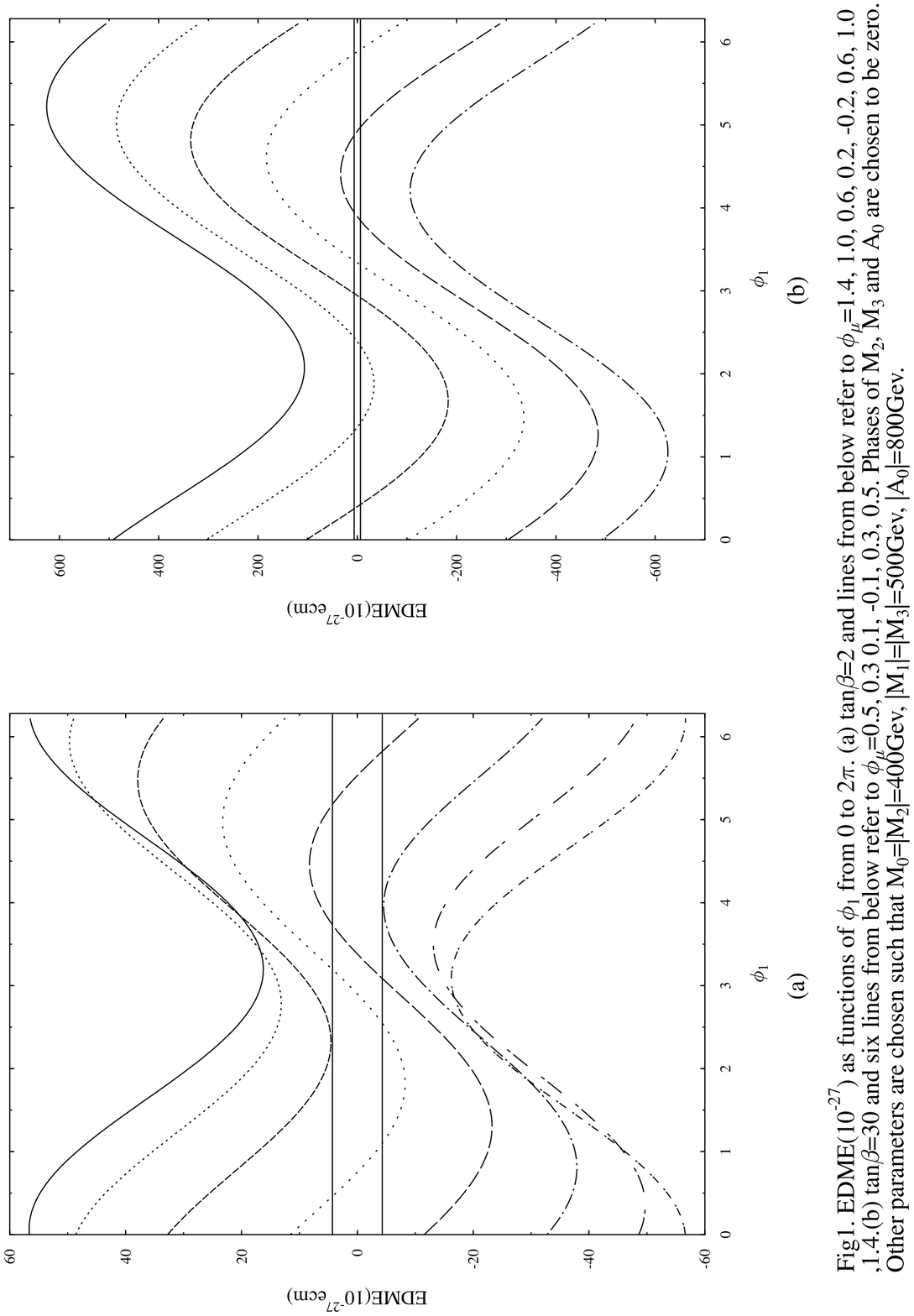,width=7.2 in}
}
\label{fig1}
\end{figure}

\newpage
\begin{figure}[t]
\vspace{-2cm}
\centerline{
\epsfig{figure=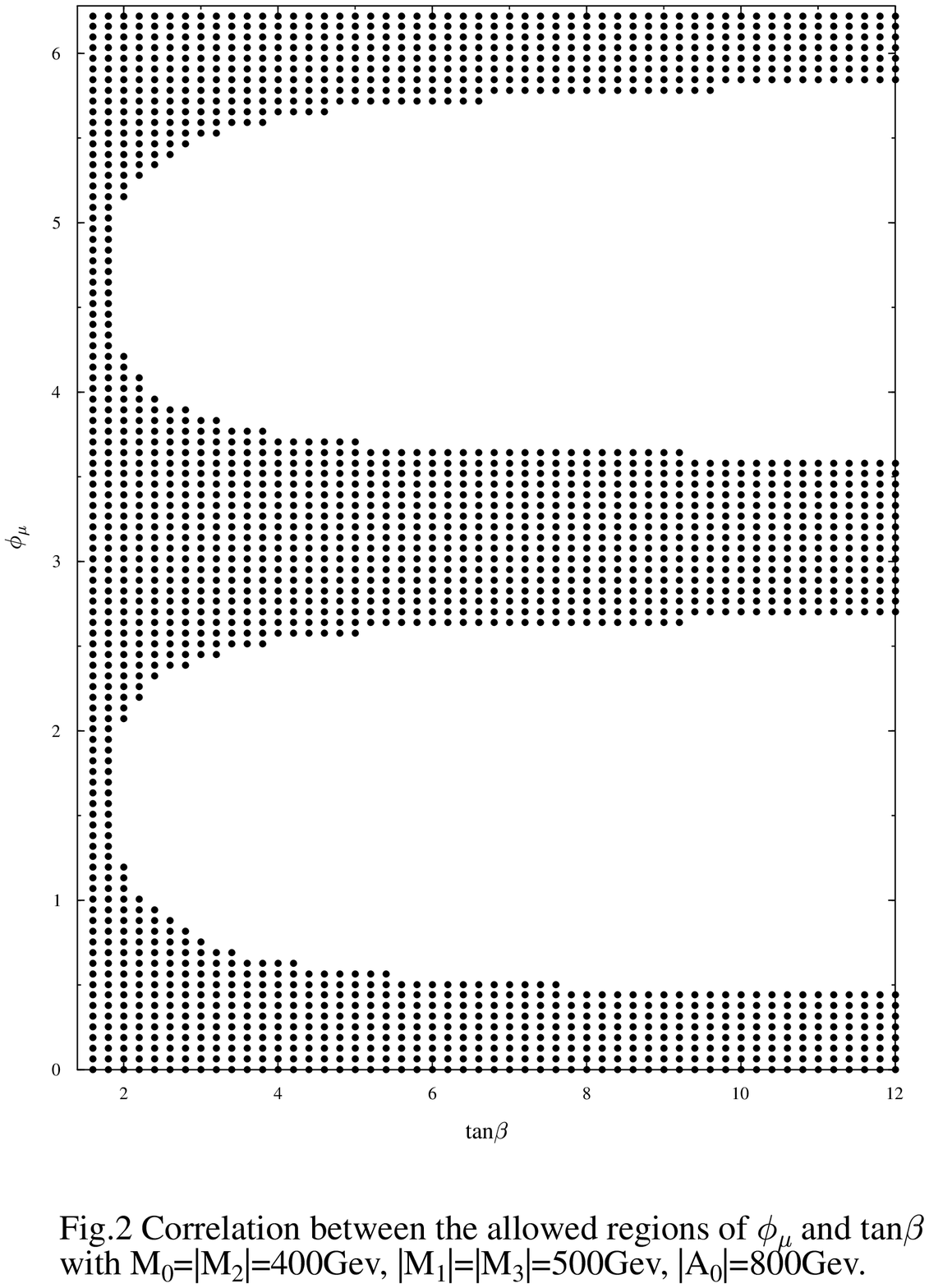,width=7.2 in}
}
\label{fig2}
\end{figure}

\newpage
\begin{figure}[t]
\vspace{-2cm}
\centerline{
\epsfig{figure=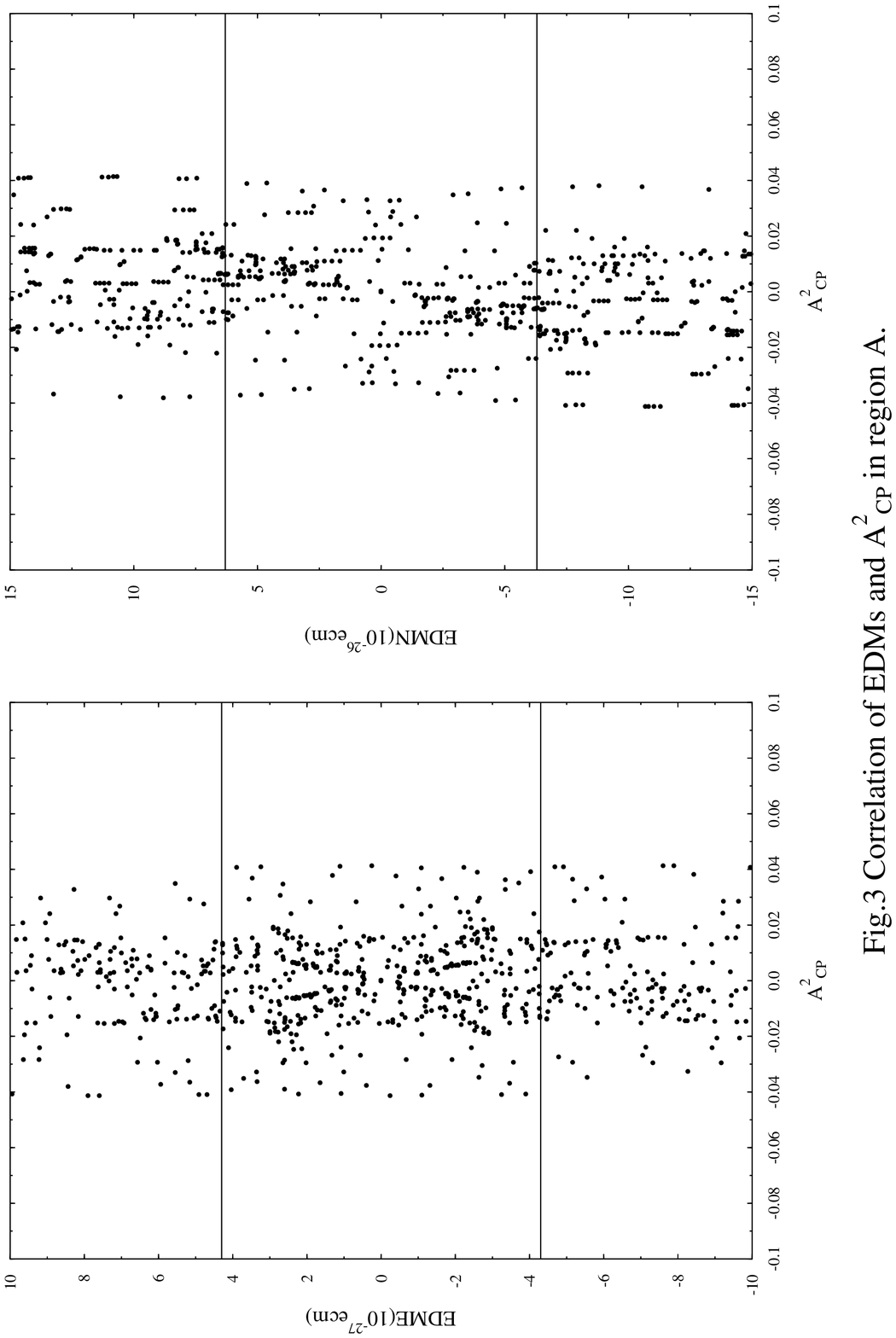,width=7.2 in}
}
\label{fig3}
\end{figure}

\newpage
\begin{figure}[t]
\vspace{-2cm}
\centerline{
\epsfig{figure=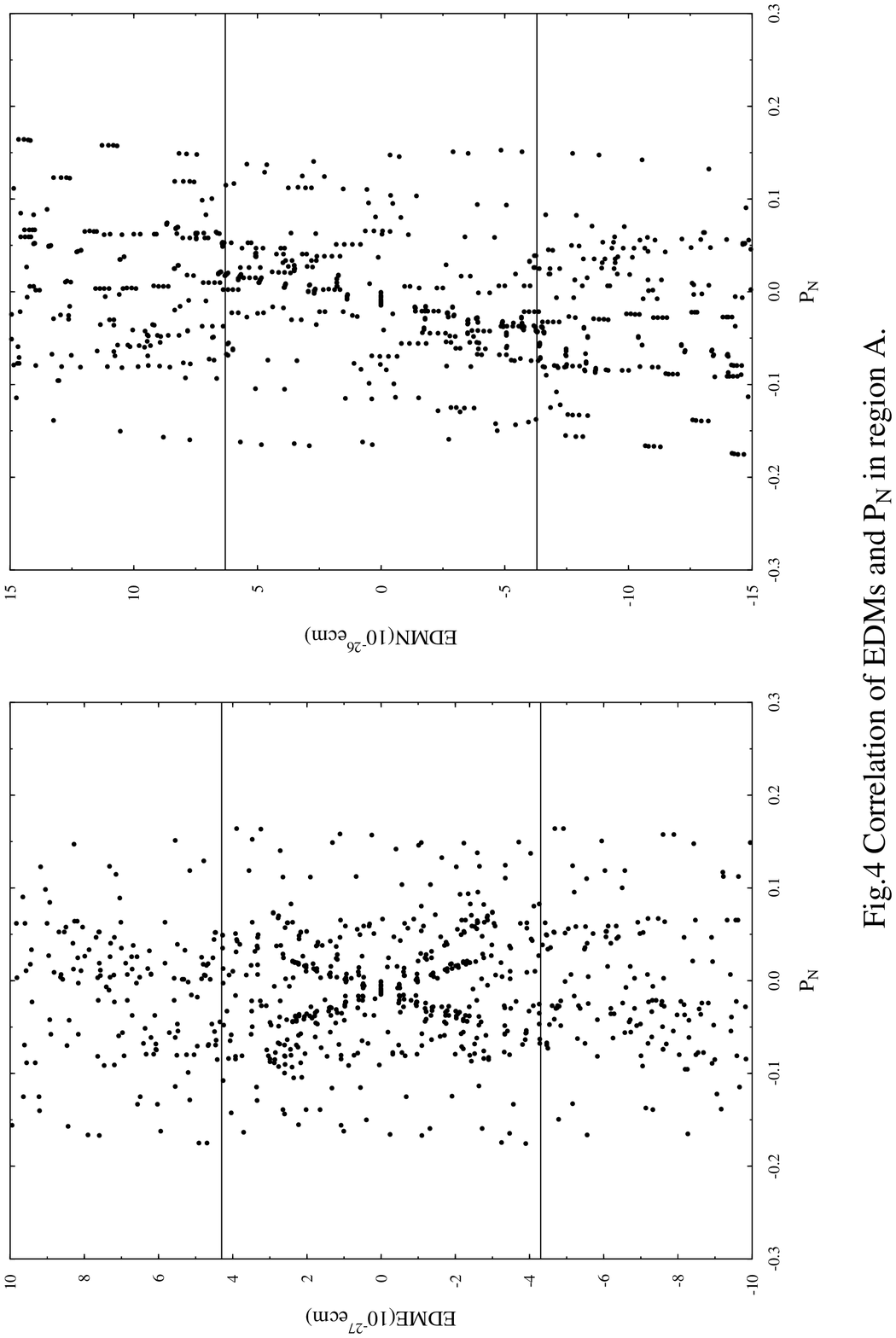,width=7.2 in}
}
\label{fig4}
\end{figure}


\begin{thebibliography}{99}
\bibitem{kt}A. Alavi-Harati et al., Phys. Rev. Lett. {\bf 83} (1999) 22.
\bibitem{na}G.D. Barr et al., NA31 collaboration, Phys. Lett. {\bf B317} (1993) 233.
\bibitem{old} M.Dugan,B.Grinstein, and L.J.Hall, Nucl.Phys. {\bf B255}
413 (1985); S.Dimopoulos and S.Thomas,Nucl.Phys. B{\bf 465},23(1996);
J. Ellis, S. Ferrara and D. V. Nanopoulos, Phys. Lett. {\bf B114} (1982) 231;
for a review see S. M. Barr and W. J. Marciano, in CP Violation, edited by C. Jarlskog (World
Scientific, Singapore, 1989), p. 455; R. Barbieri, A. Romanino, and A. Strumia, Phys. Lett.
{\bf B369} (1996) 283.
\bibitem{commins} E. Commins, et.al, Phys. Rev. {\bf A50}, 2960(1994).
\bibitem{HLG} P.G. Harris et.al, Phys. Rev. Lett.{\bf 82},904(1999);
see also S.K. Lamoreaux and R. Golub, hep-ph/9907282.
\bibitem{IN} T.Ibrahim and P.Nath, Phys. Lett. {\bf B418} 98 (1998), Phys.Rev. D{\bf 57},
478(1998), (E) ibid, {\bf D58}, 019901(1998), Phys. Rev. {\bf D58},111301(1998), hep-ph/9910553.
\bibitem{others}T. Falk and K. Olive, Phys. Lett. {\bf B439}, 71(1998); S. Pokorski, J. Rosiek,
and C. Savoy, hep-ph/9906206. 
\bibitem{HL} Chao-Shang Huang and Liao Wei, hep-ph/9908246.
\bibitem{bgk}M.Brhlik,G.J.Good,and G.L.Kane, Phys. Rev. {\bf D59},11504(1999).
\bibitem{KSAH} A. Ali and G. Hiller, Eur.Phys.J. {\bf C8}(1999) 619-629;
 F. Kruger and L.M. Sehgal, Phys.Rev.{\bf D55}(1997) 2799.
\bibitem{bim}A. Brignole, L. Ib$\acute{a}\acute{n}$ez, C. Mu$\acute{n}$oz, Nucl. Phys. {\bf B422}
(1994) 125, Erratum-ibid Nucl. Phys. {\bf B436} (1995) 747.
\bibitem{rge}K. Inoue et al., Prog. Theor. Phys. 68 (1982) 927; A. Bouquet, J. Kaplan and C.A. 
Savoy, Nucl. Phys. B 262 (1985) 299; V. Barger, M. Berger and P. Ohmann, Phys. Rev. {\bf D49} 
(1994) 4908.
\bibitem{ckp}D. Chang, W.-Y. Keung, A. Pilaftsis, Phys. Rev. Lett. {\bf 82} (1999) 900.
\bibitem{hz}C.-S. Huang and S.-H. Zhu, Phys. Rev. {\bf D61} (2000) 015011.
\bibitem{Lee}T.D. Lee and C.S. Wu, Annu. Rev. Nucl. Sci. {\bf 16} (1966) 471.
\bibitem{Rai} S.Rai Choudhury et al., hep-ph/9902355, where $P_N$
have been given,but they gave only two terms in the numerator of $P_N$.
\bibitem{hy}C.S. Huang and Q.S. Yan, Phys. Lett. {\bf B442} (1998) 209; C.S. Huang, L. Wei, and Q.S. Yan, Phys. Rev.
{\bf D59} (1999) 011701.
\bibitem{bb}S. Bertolini, F. borzumati, A. Masieso and G. Ridolfi, Nucl. Phys. B 353 (1991) 591;
P. Cho, M. Misiak and D. Wlyer, Phys. Rev. D 54 (1996) 3329.
\bibitem{CKKM} S. Codoban and D.I.Kazakov, hep-ph/9906256; D. Kazakov AND G. Moultaka, hep-ph/9912271
\bibitem{aad}E. Accomando, R. Arnowitt, and B. Dutta, hep-ph/9907446.
\bibitem{cleo} S. Ahmed et.al, CLEO collaboration, CLEO CONF 99-10,
hep-ex/9908022
\bibitem{goto}T. Goto et al.,Phys.Lett. B460 (1999) 333-340.
\bibitem{ls} E. Lunghi and I. Scimemi, hep-ph/9912430.
\end{thebibliography}
\end{document}